\documentclass{JHEP3}

\usepackage{epsfig}

\newcommand{\be}{\begin{equation}}
\newcommand{\ee}{\end{equation}}

\title{Infinite $N$ phase transitions in continuum Wilson loop operators.}
\author{R. Narayanan
\\Department of Physics, Florida International University, Miami,
FL 33199, USA\\E-mail: \email{rajamani.narayanan@fiu.edu}}
\author{ H. Neuberger
\\ Rutgers University, Department of Physics and Astronomy,
Piscataway, NJ 08855, USA\\E-mail: \email
{neuberg@physics.rutgers.edu} }

\abstract {We define smoothed Wilson loop operators on a four
dimensional lattice and check numerically that they have a finite
and nontrivial continuum limit. The continuum operators maintain
their character as unitary matrices and undergo a phase transition
at infinite $N$ reflected by the eigenvalue distribution closing a
gap in its spectrum when the defining smooth loop is dilated from
a small size to a large one. If this large $N$ phase transition
belongs to a solvable universality class one might be able to
calculate analytically the string tension in terms of the
perturbative $\Lambda$-parameter.
This would be achieved by matching instanton results for
small loops to the relevant large-$N$-universal function
which, in turn, would be matched for large loops to
an effective string theory.
Similarities between our findings and known analytical results in
two dimensional space-time indicate that the phase transitions we
found only affect the eigenvalue distribution, but the traces of
finite powers of the Wilson loop operators stay smooth under
scaling.}


\preprint{}

\begin{document}

\section{Introduction.}

Most of the concrete results of this paper were obtained by methods of
lattice field theory. These results make up a first component of a
larger, rather ambitious and quite speculative
program which is defined in the continuum, independently of any
regularization. The outline of the program makes up the bulk of
this section. The intention is to put the presentation of the lattice
results that follows in later sections in perspective. Throughout
the paper, we shall get back to continuum language to emphasize
what structures are, in our view, continuum properties.

Pure four dimensional $SU(N)$ gauge theories are strongly
interacting at large distances and weakly interacting at short
scales. In $SU(3)$ the cross-over between these two regimes is
relatively narrow and there are no good analytic approximation
methods that apply to it; nor do we have a continuum way to
calculate at strong coupling. The cross-over narrows as the number
of colors, $N$, increases.

Here, in a direct continuation of the line of thought of some old
work \cite{old1, gw12}, we report on an improved line of attack on this
problem and take several steps in its direction. The old idea was
that instantons indicate that in the limit of infinite $N$, at a
particular sharply defined scale, there is a phase transition, and
weak and strong coupling physics can be connected at that point.
This idea has not been successfully implemented because too many
parts were either wrong or missing. Much has been learned in the
interim period and we would like to try again. Let us now briefly
overview the new ingredients of our approach:

While the guess that the cross-over regime
collapses to a point and becomes a phase transition at
infinite $N$ is old, we
understand now that the scale at which the transition occurs is
not a universal quantity that applies to all observables
simultaneously; nor does it necessarily affect any local
observables: it is not seen in the free energy density for
example.
The most interesting large $N$ transitions do not have
to be ``bulk'' phase transitions.

Recent lattice work on large $N$ has produced such ``bulk''
transitions at global
critical scales in the planar limit of QCD on a Euclidean torus,
but the associated transitions take the system out of the zero
temperature, infinite volume regime and depend on the global
topology of space-time. This is not going to directly explain how
a few particle high energy process hadronizes, for example.

Recent lattice work has also indicated that
specific non-local observables undergo large
$N$ phase transitions as they are dilated \cite{bulk2,knn} and this
happens while the Euclidean space-time volume stays infinite in
all directions. The critical scales are observable dependent; one
can imagine associating with these observables an entire family of
dilated images and somewhere along the dilation axis the large $N$
transition takes place.

If there were a string equivalent to
planar QCD, one would guess that the dilation axis is somehow
related to a fifth coordinate as has been often hypothesized
\cite{poly3,malda4}. A meaningful higher dimensional formulation
presupposes a local definition of the theory in the extended
space; we shall have something to add to this speculation at the
end of the paper, but it will be vague.

Another new ingredient we add to the old program is the modern
understanding of universal features of certain large $N$
transitions in matrix models. Although, on the face of it, there
is no concept of locality in the index space associated with the
symmetry group, the models admit a
't Hooft double line expansion for two-index representations
classified by the genus of two
dimensional surfaces. A concept of two-dimensional locality
applies to the physics on these surfaces and the associated
Liouville type models represent universality classes of finite,
but varying degrees of stability associated with the large $N$
transitions \cite{doublesca5}. Most of these models are exactly
solvable.

We hope that the transitions we shall exhibit belong to some such
universality class. We have a guess for the relevant universality
class and it has an exactly solvable matrix model representative.
This means that at finite but large enough
$N$, when the transition dissolves into a cross-over, there is a
description of well chosen observables undergoing the cross-over
in terms of universal scaling functions with a dependence on a
limited number of non-universal parameters that typically come in
as boundary conditions on the equations determining the universal,
known scaling functions.

It is through these parameters that the
cross-over regime connects to the weak and strong coupling regimes
on either side. The four dimensional dilation parameter, in the
vicinity of its critical value plays the role of a
coupling that turns critical in the relevant version of matrix model
description of the large $N$ transition. By the mechanism of a
(double) scaling limit the dilation parameter
gets connected to a scale in
the two dimensional physics that codifies the universal features
of the large $N$ transition. This connection will involve a
nontrivial power and an associated dimensionful parameter.

The weak and strong coupling regimes are governed by their own
universal behavior, field theoretical on the weak side, and (we
guess) effective string like on the strong coupling side. Thus,
the free, nonuniversal, parameters needed by
each universal description, field
theoretic, large $N$ and string-like are matched to each other as
the operator is dilated, and several universality mechanisms
coexist in the same system, controlling separate scale ranges.

If the above scenario holds at $N=\infty$, we can apply it to pure
$SU(3)$ gauge theory, and, hopefully provide a
calculation of the string tension in terms of $\Lambda_{\rm QCD}$
which is accurate to order $\frac{1}{ N^2}\approx 10$ percent. For
a physicist, this might be of more value even  than a mathematical
proof of confinement. We are far from realizing this goal.

The first hurdle to overcome is to find good observables; we claim
that the right question is not whether or not there is a large $N$
phase transition in general, but rather, can one find good
observables which undergo such a transition and can be controlled
(by different theories) at very short and at very long distance
scales respectively. Local observables are unlikely to undergo the
transition. Therefore the cherished analyticity properties in
four dimensional Euclidean momentum space stay protected,
maintaining the basic unitarity and covariance properties of the
field theoretical description of the physics.

In this paper we define a
class of smeared Wilson loop observables on the lattice with the
intent that they have finite continuum limits and obey the
following properties:
\begin{itemize}
\item It is correct to think about the parallel
transport round the loop as being realized by an $SU(N)$ matrix.
This statement has to hold after the renormalization of the theory
and of the operator have been carried out.
\item When we scale the loop, a phase
transition takes place in its eigenvalue distribution if
$N=\infty$.
\item The smeared loop operator is calculable in
perturbation theory when it is scaled down.
\item Its shape dependence
is well described by an effective string theory when it is scaled
up.
\item It is possible to see by continuum non-perturbative
methods (e. g. instantons) that the phase transition is approached
when the loop is dilated up from being very small, so one can hope
to calculate the matching at the weak coupling end of the
cross-over.
\end{itemize}
At the large scale end one also needs a way to
calculate a matching, and this is presumed to work by the large
scale end of the cross-over having some string description (coming
from the relevant universal matrix model) which can be joined
smoothly to the right effective four dimensional effective
string theory. That effective string theory
could be, for example, similar
to \cite{ps6}. This we have not yet established, and is left for
future work. Nevertheless, on the basis of the results
we do have, we believe to have found
non-local observables which are ``good'' enough.

In its essence the calculation we have
outlined is an attempt to extend the
logic of effective field theory \cite{effft7} calculations to the
strong-weak cross-over in pure gauge QCD. Also, the program is in
the same spirit as the old CDG approach to instantons
\cite{cdg75} where they tried to use instanton
gas methods to set the parameters in an effective bag-model
description of hadrons.

If our scenario works for pure glue, it
would be possible to re-introduce fermions at a later stage, at
least as probes with no back reaction and see whether one can
construct hadronic observables containing fermionic fields that
also undergo large $N$ transitions. In particular it would be
interesting to find a
similar calculational program
for spontaneous chiral symmetry breaking.
One could check this numerically using exactly chiral
fermions~\cite{overlap}.

In the next section we introduce our observables on the lattice
and present our numerical results in support of the claim that our
observables have a continuum limit and undergo large $N$
transitions under dilation. In subsequent sections we discuss a
continuum view of the lattice construction and our guess for the
large $N$ universality class. The presentation
becomes by necessity progressively more speculative. Ultimately,
we hope to reduce the role of lattice simulations to a strictly
qualitative one, providing support for the assumptions about the
various universality classes that control the different regimes
our calculation needs to connect.
The calculation in itself would be entirely
analytical. It goes without saying that any analytical step would
need to be checked against numerical work, as this is unchartered
territory.

\section{Overview of lattice study.}

We use an ordinary single plaquette Wilson action defined on a
hypercubic lattice. We exploit twisted reduction \cite{twist8}
which allows a decent interval of lattice spacings to be realized
on a small lattice of size $V=7^4$.
For twisting to work
best, we need $N$ to be given by $4$ times a prime number.
We work mostly at $N=44$ with
checks for finite $N$ effects at $N=28$.
Our simulation method
employs a combination of heat-bath and over-relaxation updates and
equilibration is achieved in reasonable lengths of computer time.
In most of our analysis we can ignore all statistical errors.
As with any lattice work with few
precedents to rely on, a substantial part of the effort is to
identify useful ranges of the many parameters that enter numerical
work, so that the relevant physics questions can be addressed.
Throughout, we use $b$ for the lattice gauge coupling which is the
inverse bare 't Hooft coupling; relatively to the conventional
lattice coupling $\beta$, $b$ already has the right power of $N$
extracted:
\be
b=\frac{\beta}{2N^2}=\frac{1}{g_{\rm YM}^2N}\ee
For clarity purposes, we ignore a technicality in our
presentation:
For a twisted simulation $b$ is taken as the
negative of its value in the untwisted case; we shall quote the
conventional, positive value of $b$.

\subsection{Smeared lattice Wilson loops.}

We consider only rectangular loops restricted to one plane of our
hypercubic lattice and most of our work is done with square loops.
However, a continuum approach would
be concerned with the class of
observables that contain smooth loops of arbitrary shape.
Specifically, explicit calculations in the continuum would best be applied to
circular loops contained in one flat plane in $R^4$, as they have
maximal symmetry.

In order to eliminate the perimeter and corner
divergences \cite{div9} we construct the Wilson loop operators
associated with our curves not from the $SU(N)$ link matrices
$U_\mu (x)$ generated by the probability distribution determined
by the single plaquette Wilson action, but, rather, from smeared
versions of those. We employ APE smearing \cite{ape10}, defined
recursively from $n=0$, where the smeared matrix is equal to
$U_\mu(x)$. Let $\Sigma_{U^{(n)}_\mu (x;f)}$ denote the ``staple''
associated with the link $U^{(n)}_\mu(x;f)$ in terms of the entire
set of $U^{(n)}_\nu(y;f)$ matrices. One step in the recursion
takes one from a set $U^{(n)}_\mu (x;f)$ to a set $U^{(n+1)}_\mu
(x;f)$:
\begin{eqnarray}
X^{(n+1)}_\mu (x;f)= (1-|f|) U^{(n)}_\mu (x;f)+\frac{f}{6}
\Sigma_{U^{(n)}_\mu (x;f)}\nonumber\\ U^{(n+1)}_\mu (x; f
)=X^{(n+1)}_\mu (x;f) \frac{1}{\sqrt{[X^{(n+1)}_\mu (x;f)]^\dagger
X^{(n+1)}_\mu (x;f)}}
\end{eqnarray}
In the simulation, one never encounters a situation where the
unitary projection in the above equations stalls because $X^{(n)}$
is singular. In other words, smearing is well defined with
probability one.
For a twisted simulation $f$ is taken as the
negative of its value in the untwisted case; we shall quote the
conventional, positive value of $f$.

$U^{(n)}_\mu (x;f)$ transforms under gauge transformations the
same way as $U_\mu (x)$ does. Our smeared Wilson loop operators,
$\hat W [L_1,l_2 ; f ; n]$ are defined as ordered products round
the $L_1 \times L_2$ rectangle restricted to a plane. $L_\alpha$
are integers and give the size of the loop in units of the lattice
spacing. When the traversed link starts at site
$x=(x_1,x_2,x_3,x_4)$ $ x_\mu\in Z$ and connects to the
neighboring site in the positive direction $\mu$, $x+\mu$, the
link matrix is $U^{(n)}(x;f)$, while when this oriented link is
traversed in the opposite direction, the link matrix is
$U^{\dagger (n)}(x;f)$. $\hat W$ depends on the place the loop was
opened, but its eigenvalues do not. The set of eigenvalues is
gauge invariant under the fundamental gauge transformation
operating on $U_\mu (x)$.

We established by numerical means, to be described in greater
detail later, that as $L_1 , L_2 , f, n$ are varied, at specific
lattice couplings, the spectrum of $\hat W [L_1 , L_2 ,; f ; n] $
opens a gap for very small and/or very smeared loops. This gap
closes for very large and/or very lightly smeared loops.

\subsection{Lattice large $N$ transitions survive in the continuum limit.}

Even restricting to just square loops, our lattice operators
depend on many parameters: there is an overall scale, the side of
the square, $L$; there is a gauge coupling $b$ which fixes the
lattice spacing in physical units, and therefore also controls the
scale; there are two parameters, $f$ and $n$ that control the APE
smearing and are responsible of the taming of the perimeter and
corner divergences in the traces of the operator in all
irreducible $SU(N)$ representations simultaneously. Our
construction implicitly adjusts the infinite number of
undetermined finite parts associated with the renormalization of
the trace in each representation so that the entire collection can
be viewed as coming from one (fluctuating) $SU(N)$ matrix. This
ensures that backtracking segments of the loop make no
contribution to the operator.

Obviously, there is a substantial amount of arbitrariness in the
choices we made when defining the regularized Wilson loop
operators. If the construction is kept more general, this
arbitrariness would reflect itself in an invariance under a large
set of transformations in the continuum, perhaps under the form of
a generalized renormalzation group equation.

Our first objective is to find out how to adjust the parameter
dependence in $\hat W$ such that $N=\infty$ transition points
identified on the lattice, where the gap just opens, survive in
the continuum limit in which the lattice coupling $b$ is taken to
infinity together with $L_{1,2}$ in such a way that the physical
lengths $l_\alpha = L_\alpha a(b)$ are kept fixed. $a(b)$ is the
lattice spacing at coupling $b$, and we express it in terms of the
physical finite temperature deconfining transition $t_c$, by
\be
a(b)=\frac{T_c(b)}{t_c}
\ee
where $\frac{1}{T_c(b)}$ is the lattice
size of the compact direction on an $S^1\times R^3$ lattice
corresponding to the finite temperature transition critical
coupling $b$.

We set the number of smearing steps $n$ to be proportional to the
perimeter square (we restricted the loop sizes to even
$L_1+L_2$), $n=\frac{(L_1+L_2)^2}{4}$. The physical
sizes of the loop are $l_\alpha$ and $L_\alpha = \frac{l_\alpha}{a(b)}$. 
We have
set $n=\frac{(L_1 (b) + L_2 (b))^2}{4}$ because in physical terms
the product $f n$ is a length squared. This can be seen from the
simple perturbative argument to be presented in
the next subsection, or, more
intuitively, by observing that smearing is a random walk that
fattens the loop and the thickness grows as the square root of the
number of smearing steps. Our choice for $n$ makes $f$ a
dimensionless parameter in the physical sense; on the lattice $f$
is actually bounded to an interval of order one.

We analyzed four sequences of $\hat W [L_1 , L_2; f;
n=\frac{(L_1 + L_2 )^2}{4}]$ keeping the physical size fixed at
four values, one for each sequence. We set $L_1=L_2=L$.
By various means, to be described later, we varied $f$ and found the
critical value of $f$ at which each member in each sequence
just opens a gap in its spectrum at eigenvalue $-1$.
(So long as CP invariance is maintained, for any loop, 
the eigenvalue probability is
extremal at $\pm 1$.) The sets of critical values for 
each sequence were then extrapolated
to the $b\to\infty$ limit. In each sequence
the physical size of the loop is kept fixed by definition. This gave four
continuum critical values of $f$ for the four different
physical loop sizes we selected. Throughout, the values of $f$ used were within
reasonable ranges, well within the maximally allowed lattice range
of $0<f<0.75$ \cite{pertsmear11}.
That it is actually possible to arrange that only such reasonable values of $f$
are needed in the vicinity of the large $N$ transitions
is our most basic result and a nontrivial feature of the dynamics
of planar pure Yang Mills theory. The collection of the continuum
limits of the critical values of $f$ at the
transition points associated with each sequence identified by its
physical scale defines a critical line, $f_c(l)$.
For any finite $b$ there were finite lattice spacing effects, but they
were sufficiently small to keep all the lattice approximations to
$f$ in a good range.

Thus, we conclude that we have found a continuum critical line
$f_c(l)$ in planar QCD where our continuum Wilson loop operators
have a well defined spectrum and that spectrum is critical in the
sense that its eigenvalue density, $\hat \rho (\theta)$ is nonzero
for any $\theta\ne\pi$ but $\hat\rho (\pi)=0$. Here, we denote the
eigenvalues of the Wilson loop operator as $e^{\imath\theta}$,
where $\theta$ is an angle on a circle.

It is not necessary to keep $n$ quadratic in $L_\alpha$ for
$\Lambda l_\alpha >>1$; the loops that have a critical $f$ are of
sizes $l$ for which $\Lambda l $ is of order unity. For much
larger loops, $f n$ should be allowed to become proportional to a
physical scale of order $\frac{1}{\Lambda^2}$ and not increase
further as the physical loop size goes to infinity. All that is
important is that for loop sizes of order $\frac{1}{\Lambda}$, $n$
be adjusted as above, and that for small loops we let $f n$ become
small too. For small loops we wish to enter a perturbative regime,
and therefore we should not eliminate the ultraviolet fluctuations
at short distances, where they dominate.

Once the existence of the $f_c (l)$ line has been established in
the continuum limit, with $\frac{df_c}{dl} > 0$, we can be
reasonably sure that dilating a $\hat W [L,L;f;n]$ so that at
lattice coupling $b$ it passes through a point $L=\frac{l}{a(b)},
f=f_c (l)+{\cal O}(a^2(b)), n=L^2$ the spectrum of this $\hat W$
will go through the transition. We still need to make sure that
once the gap is closed, the density at $-1$ starts building up
continuously from zero; this would mean that the transition is
continuous and likely to possess a universal character. If the
transition were discontinuous, like, for example, the finite
temperature transition in Polyakov loop operators, it is unlikely
that a scaling regime exists on either side of the transition and
a single scaling regime connecting the two sides (of the type we
are hoping for) is ruled out. This is obviously the case in the
finite temperature case because the transition there is not
induced by taking $N\to\infty$, but exists for all $N$, and starts
being first order at $N=3$.

We addressed this point by carrying out additional simulations
along another line in the physical $(f,l)$ plane, which is not a
constant $l$ but, rather, has $l$ varying by a substantial amount.
We picked a convenient interval of lengths $l$ and a line in the
plane $(f,l)$ given by $f=f_0-f_1 l^2$ with constant $f_1 >0$
chosen so that $f$ varies between $f_1$ and $f_2$ with $f_1\approx
0.1$ and $f_2\approx 0.2$. This line intersects $f_c (l)$ at an
angle not too close to zero, thus enhancing the variability in the
spectrum as one takes $\hat W$ along the line between the two
extremal lengths. There is no particular physical meaning attached
to this second line, except that along it the loop is dilated and
in addition, the amount of smearing decreased; both effects help
closing the spectral gap that is present when the loop is small.

To go along the continuum line we fixed the lattice loop-size $L$
and varied $b$. We kept $n=L^2$ also fixed, but
varied $f$ according to the formula $f=f(b)=f_0-f_1 (La(b))^2$. 
Following $\hat W$
along this line we established that indeed the spectrum of $\hat
W$ opens a gap as the band of lattice curves 
converging to $f_c(l)$ is crossed.

We avoided working close to the transition and estimated
independently on each side where the corresponding phase would
end, using various extrapolation methods: On the side where there
is a gap, we studied the decrease of the gap to estimate when it
will close. On the gap-less side, we estimated when $\hat \rho
(\pi)$ would vanish.

We repeated the procedure for three increasing values of $L$.
In each case one covers similar (overlapping, but not identical)
ranges of physical size $l$. In the common portions of the range,
the larger the lattice value of $L$ is, the smaller the lattice
effects ought to be. We found general consistency with
a scaling violation that goes as the lattice spacing squared.

The separate extrapolations in each phase produced independent
estimates for end points of each phase which came out
reasonably close to each other. This is consistent with a continuous large
$N$ transition at which the eigenvalue distribution of $\hat W$
opens a gap. The determinations of the critical points from each
side were ordered in a way that was inconsistent with a
discontinuous transition in which $\hat \rho(\pi)$ decreases to a
non-zero value and then drops to zero, jumping into a phase with a finite
gap around $\pi$. Thus, we conclude that the transition is
continuous and one can hope for some universal
description in the vicinity of the transition point. Actually, it
is hard to imagine what kind of dynamics could have 
produced a first order large
$N$ transition in the spectrum.

\subsection{Smearing removes ultra-violet divergences.}

The effect of smearing is easy to understand in perturbation
theory where one supposes that each individual step in the
smearing iteration can be linearized. Writing $U^{(n)}_\mu
(x;f)=\exp(iA^{(n)}_\mu (x;f))$, and expanding in $A_\mu$ one
finds~\cite{pertsmear11}, in lattice Fourier space:
\be
A^{(n+1)}_\mu (q;f)= \sum_\nu h_{\mu\nu} (q) A^{(n)}_\nu (q;f) \ee
with
\be
h_{\mu\nu} (q)= f(q)(\delta_{\mu\nu} - \frac {{\tilde q}_\mu
{\tilde q}_\nu}{\tilde q^2})+\frac {{\tilde q}_\mu {\tilde
q}_\nu}{\tilde q^2} \ee where $\tilde q_\mu = 2\sin
(\frac{q_\mu}{2})$ and
\be
f(q)=1-\frac{f}{6}\tilde q^2
\ee
The iteration is solved by
replacing $f(q)$ by $f^n (q)$, where, for small enough $f$,
\be
f^n(q) \sim e^{-\frac{f n}{6} \tilde q^2}
\ee

The condition $0<f(q)<1$ for all momenta, requires $0<f<0.75$. To
get some feel for the numbers, a $4\times 4$ loop, at a value of
$b$ that makes it critical, will have $n=16$ and $f\sim 0.15$,
which makes $\frac{f n}{6}\sim 0.4$, which is an amount of
fattening over a distance as large as the loop itself. This is why
much larger loops should not be smeared with an $f n$ factor that
keeps on growing as the perimeter squared; rather, for a square
loop of side $L$, for example, the following choice would be
appropriate:
\be
f=\frac{\tilde f}{1+M^2 L^2}
\ee
Here, $M$ is a physical mass $m$
expressed in lattice units, $M=m a(b)$. $m$ is a typical hadronic
scale chosen so that at the large $N$ transition we found, $ML$ is
less then $0.01$, say.

To reassure ourselves that such an extension of the cutoff
dependence of $f$ indeed would connect smoothly to large loops
which show confinement with the known value of the string tension,
we roughly estimated the string tension from our smeared loops,
at the large size end of the ranges we studied. We found that the
string tension comes out close to that of bare lattice Wilson
loops, and this shows that our smeared Wilson loop operators are
acceptable pure $SU(N)$ gauge theory string candidates.

In summary, the effect of smearing on the propagator of
$A^{(n)}_\mu (x;f)$ is to eliminate the ultraviolet singularity
that gives rise to perimeter and corner divergences in traces of
bare Wilson loop operators. This works when the product $fn$
varies as a physical length squared when the continuum limit is
approached. For small loops we choose this physical length to be
proportional to the loop itself. For large loops the physical
length stabilizes at some hadronic scale. In this way, our
operators become sensitive to the short distance dynamics when the
scale is very small, but, when confinement becomes the dominating
effect, for very large loops, the string tension of the smeared
Wilson loops is the true, universal, string tension of the theory,
unaffected by the smearing. The leading term in the effective
string theory, describing the shape dependence of the Wilson loops
when the loop is very large, is expected to be universal and the
string tension enters only in a trivial manner, as a dimensional
unit.

We stress that we are not executing a renormalization group
transformation; our Wilson loop operators are defined for
infinitely thin loops in space-time and for all loop sizes;
rather, we are applying a regularization whose objective is to
make the Wilson loop operators finite in the continuum. If we
dealt with each irreducible representation separately, the
elimination of perimeter and corner divergences would leave behind
an infinite number of arbitrary finite parts. This freedom has
been implicitly exploited by imposing the requirement of unitarity
of the Wilson loop operators, which ensures what Polyakov has
termed the ``zigzag'' symmetry \cite{poly3}. Preserving the
unitary character of the Wilson loop operator in turn makes the
concept of an eigenvalue distribution well defined in the
continuum and sets the stage for possible large $N$ phase
transitions. The zigzag symmetry also influences the form of the
lattice loop equations at infinite $N$, which are obeyed by traces
of bare Wilson loops in the fundamental representation; the
effective string theory for large loops needs not obey the zigzag
symmetry, but this symmetry could provide a useful constraint on
its couplings.

\section{Numerical details of lattice study.}

In the previous section we described what our strategy was to
address the question of existence of a continuous limit for the
operators and the large $N$ transition point. Also, we made an
effort to convince ourselves that the large $N$ transitions indeed
are continuous. We now proceed to describe the exact details of
our simulations and the numbers that came out.

\subsection{The determination of $f_c(l)$.}

\subsubsection{Estimating the  gap at infinite $N$.}

At infinite $N$, small $\hat W$ loops will have a spectral gap on
the lattice. We need a method to extract that gap from numerical
data at finite $N$. As we dilate the loop, and/or reduce the
amount of smearing, the gap shrinks. We need a way to extrapolate
to the point where the gap vanishes. We wish to keep away from
measurements for too small gaps, and enter that regime only by
extrapolation.

Getting the infinite $N$ value of the gap is difficult because
subleading effects are large at the gap's edge. It is reasonable
to assume that the universal behavior at the gap's edge is the
same as at the edge of a gap in simple hermitian matrix models.
Under this assumption, we have separate predictions for the
behavior of the eigenvalue density and for the extremal
eigenvalue. One can check numerically how well the universal forms
work and we find them to work quite well. To fit onto the
universal behavior we need to adjust a parameter that corresponds
to the gap size at infinite $N$. This provides the determination
of the gap sizes we use subsequently.

It is only on the side of the transition where the eigenvalue density
has a gap that
there is a regime where large $N$ corrections are dominated by
universal functions; therefore, we are able to get good estimates
for the eigenvalue gap size at infinite $N$ from data obtained at one single,
but large $N$. For good measure, we checked that the predicted rational
power dependence on $N$ holds in the regimes in which we did our
analysis. A comparison with some simulations at $N=28$ convinced
us that the power dependence on $N$ is already at its known 
universal value at
$N=28,44$.

We studied the eigenvalue
distribution of $\hat W [L , L ; f;
n=L^2;N]$ on the lattice
for several values of $L$ and $b$ at a fixed large $N$.
Let $e^{i\theta_k}, k=1,\cdots,N$ be the $N$ eigenvalues
of a fixed $L\times L$ loop on the lattice, $-\pi \le \theta_1 <
\theta_2 < \cdots < \theta_{N-1} < \theta_N < \pi$.
The sum of all angles is zero mod $2\pi$. On a lattice of $V$
sites there are $6V$ different $L\times L$ Wilson loops with
a fixed orientation; we computed the eigenvalues
of all such loops. In this manner, we obtain a
histogram of the distribution
$p_f(\theta;U)$ of all eigenvalues
in a fixed gauge field background $U$.
We also collected data for a histogram of the distribution
$p_m(\theta_N;U)$ for the largest eigenvalue in a fixed
gauge field background. In addition, we record
the mean, variance, skewness and kurtosis
pertaining to the largest eigenvalue:
\begin{eqnarray}
\mu_N(U) &=& \langle \theta_N \rangle_U;\cr
\sigma^2_N(U) &=& \langle \theta^2_N \rangle_U - \langle 
\theta_N \rangle^2_U;\cr\cr
S_N(U) &=&
\frac{ \langle \theta^3_N \rangle_U
-3 \langle \theta^2_N \rangle_U \langle \theta_N \rangle_U
+ 2 \langle \theta_N \rangle^3_U }{\sigma^3_N(U)};\cr\cr
K_N(U) &=&
\frac{ \langle \theta^4_N \rangle_U
-4 \langle \theta^3_N \rangle_U \langle \theta_N \rangle_U
+6 \langle \theta^2_N \rangle_U \langle \theta_N \rangle^2_U
-3 \langle \theta_N \rangle^4_U }{\sigma^4_N(U)}-3.
\end{eqnarray}

If the eigenvalue distribution of $\hat W$ has a sizable gap, it
is reasonable to ignore the influence of the two edges on each
other. Then, the universal behavior at the gap's edge ought to be
the same as that of the edge of the gap in simple Gaussian
hermitian matrix model. Under this assumption, we compared our
data with the behavior of the extremal eigenvalue~\cite{tracy}
and the eigenvalue density close to the edge~\cite{gff}.

The universal distribution of the
extremal eigenvalue, $p(s)$, is given by~\cite{tracy}
\be
p(s)=\left [ \int_s^\infty q^2(x) dx \right ] e^{-\int_s^\infty (x-s)q^2(x) dx}
\ee
with $q$ being the solution to the Painleve II equation,
\be
\frac{d^2q}{ds^2} = sq + 2 q^3;
\ee
satisfying the asymptotic condition
\be
q(s) \sim Ai(s)\ \ {\rm as}\ \ s\rightarrow\infty.
\ee
The mean, variance, skewness and kurtosis of the above
distribution are $-1.7710868074$, $0.8131947928$,
$0.2240842036$ and $0.0934480876$. In order to match
the lattice data for the distribution of $\theta_N$
to the above universal form,
we define $s$ according to
\be
\theta_N = \frac{s}{\alpha} + \pi(1-g)
\label{scalevar}
\ee
and set
\begin{eqnarray}
\alpha &=& \sqrt{\frac{0.8131947928}{\sigma^2_N}}\cr
g &=& 1-\frac{1}{\pi}[\mu_N + \frac{1.7710868074}{\alpha}]
\end{eqnarray}
$\alpha$ is a non-universal scale factor and $g$ is half the arc
length of the gap on in the eigenvalue distribution on the unit
circle, centered at $-1$. The quantities, $\sigma^2_N$ and $\mu_N$
are the averages over the gauge field ensemble of the previously
defined quantities, $\sigma^2_N(U)$ and $\mu_N(U)$. From the
lattice data for $\sigma^2_N(U)$ and $\mu_N(U)$, we obtain an
estimate for $\alpha$ and $g$ with errors extracted by the
jackknife method. Using these estimates, we compared the scaled
lattice data with $p(s)$ for fixed $f$, $b$, $N$ and $V$. One such
comparison, at $f=0.15$, $b=0.361653$, $L=4$, $N=44$ and $V=7^4$
with a gap of $0.1229(4)$, is shown in Figure~\ref{gap} and there
is good agreement. In spite of the fact that data is needed over
the entire range of $s$ to get a good estimate of the skewness and
kurtosis, we find agreement between lattice data and the universal
estimates to within $5\%$ for the skewness and to within $20\%$
for the kurtosis for a wide range of gap values. The estimate of
the skewness and kurtosis do not depend on the extracted values of
$\alpha$ and $g$.

\FIGURE[ht]{
\epsfig{file=gap.eps,  width=\textwidth}
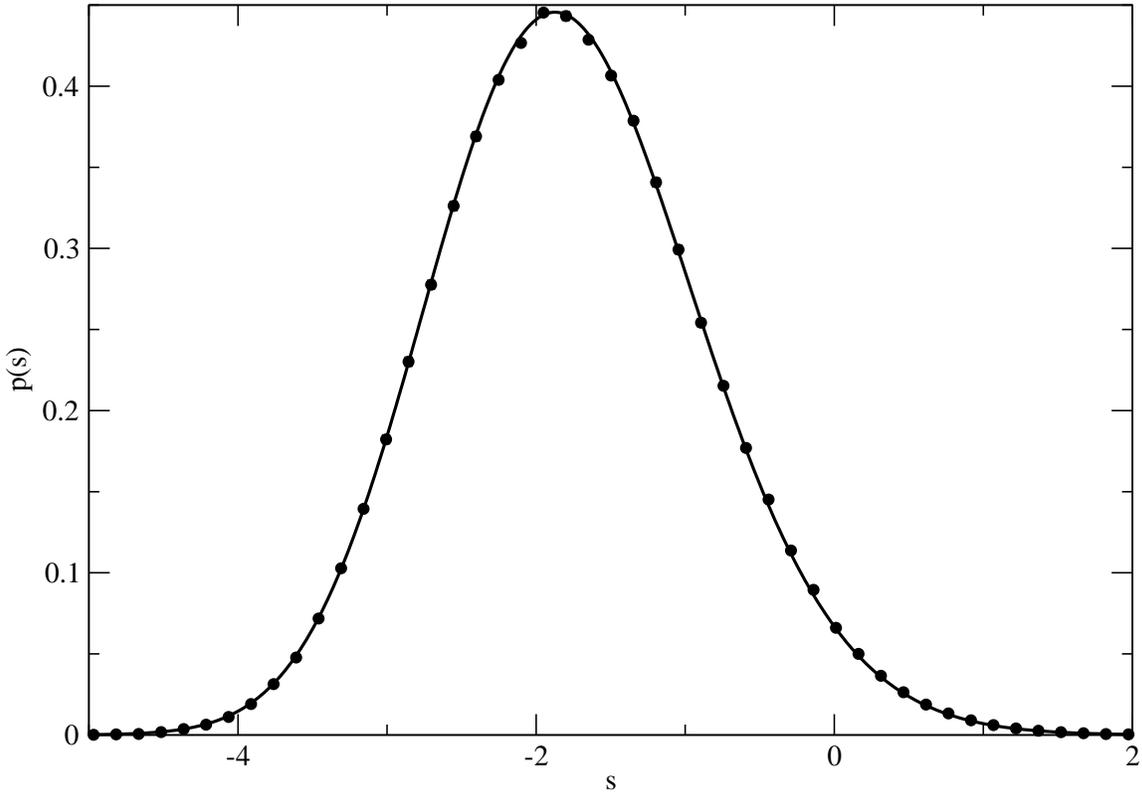
\caption{ A comparison of the
distribution of scaled extremal eigenvalue obtained on the lattice
(circles) at $f=0.15$, $b=0.361653$, $L=4$, $N=44$ and $V=7^4$
with that of the universal distribution (solid curve). }
\label{gap}}

The main result of the above numerical analysis is the
half gap size $g$, for fixed $f, L, b$, $g(f,L,b)$.

There is another way to obtain an estimate for $g$:
A recent result
provides the asymptotic corrections to the eigenvalue
density of Gaussian hermitian matrices~\cite{gff}. Using that
formula, we fitted
he eigenvalue density in terms of the scaled variable
in (\ref{scalevar}) to
\be
p_f(s) = c \left\{ [Ai^\prime(s)]^2 - s [Ai(s)]^2 \right\} +d
\left\{ 3s^2[Ai(s)]^2 - 2s [Ai^\prime(s)]^2 -3Ai(s)Ai^\prime(s)
\right\} \label{airy} \ee The variable $c$ is fixed by the overall
normalization and $\frac{d}{c}=-\frac{1}{20N^{2/3}}$ in the
asymptotic formula~\cite{gff} where $N$ is the size of the matrix.
The parameter $d$ could be non-universal; also $c$ could in some
sense be non-universal, since it depends on the total number of
eigenvalues, and it is not clear whether the Gaussian hermitian
matrix model matches onto our model with the same $N$ in both
cases. Therefore, we fitted the lattice distribution to the above
asymptotic formula with $c$ and $d$ as independent variables in
addition to the two variables appearing in (\ref{scalevar}),
producing a second determination of $g(f,L,b)$.

\FIGURE[ht]{
\epsfig{file=dist.eps,  width=\textwidth}
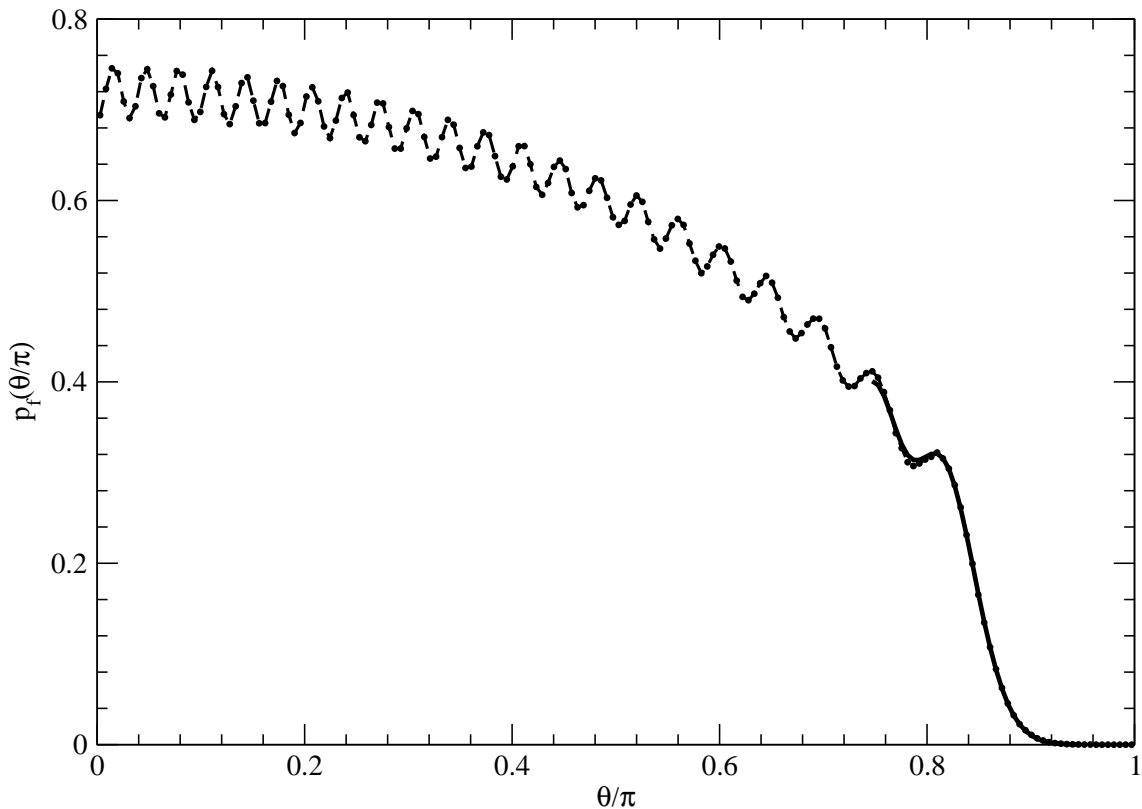
\caption{A comparison of the distribution of eigenvalue distribution
obtained on the lattice (circles with dashed lines) at $f=0.15$,
$b=0.361653$, $L=4$, $N=44$ and $V=7^4$ with that of the universal
distribution near the edge (solid curve). }
\label{dist}}

Figure~\ref{dist} shows the comparison of the full eigenvalues
distribution with the universal distribution near the edge. The
universal distribution captures the tail for the same reason that
the distribution of the extremal eigenvalue is in agreement with
$p(s)$. But we also see good agreement with the first two
oscillations near the edge and this includes more than the
extremal eigenvalue. In general we found the determination of the
gap to be consistent with the one obtained using only the extremal
eigenvalue. For an error estimate, we only used the extremal
eigenvalue method.

\subsubsection{Extrapolating to zero gap.}

The next step is to extract the critical value of $f$ at a
fixed $b$ and $L$ at infinite $N$. This is done by
assuming a fit of the form
\be
g(f,L,b) = A(L,b) \sqrt{f - F_c(L,b)}
\ee
The square root dependence on external parameters that enter the
probability law in an analytic manner is generic in matrix models
and well supported by the data.

A plot of $g^2(f,L,b)$ as function a $f$ at $b=0.361653$, $N=44$
is shown in figure~\ref{gap_vs_ff}. $F_c(L,b)$ is extracted using
a range of $f$ such that the gap, $g(f,L,b)$, is not too close to
$0$ nor is it too far from $0$. If we go to too small gaps we fear
unaccounted for rounding effects; if the gap is too large, there
is little reason to use the universal square root behavior we
postulated.

Figure~\ref{gap_vs_ff} also shows the fit along with the estimate
for $F_c(L,b)$.

\FIGURE[ht]{
\epsfig{file=gap_vs_ff.eps,  width=\textwidth}
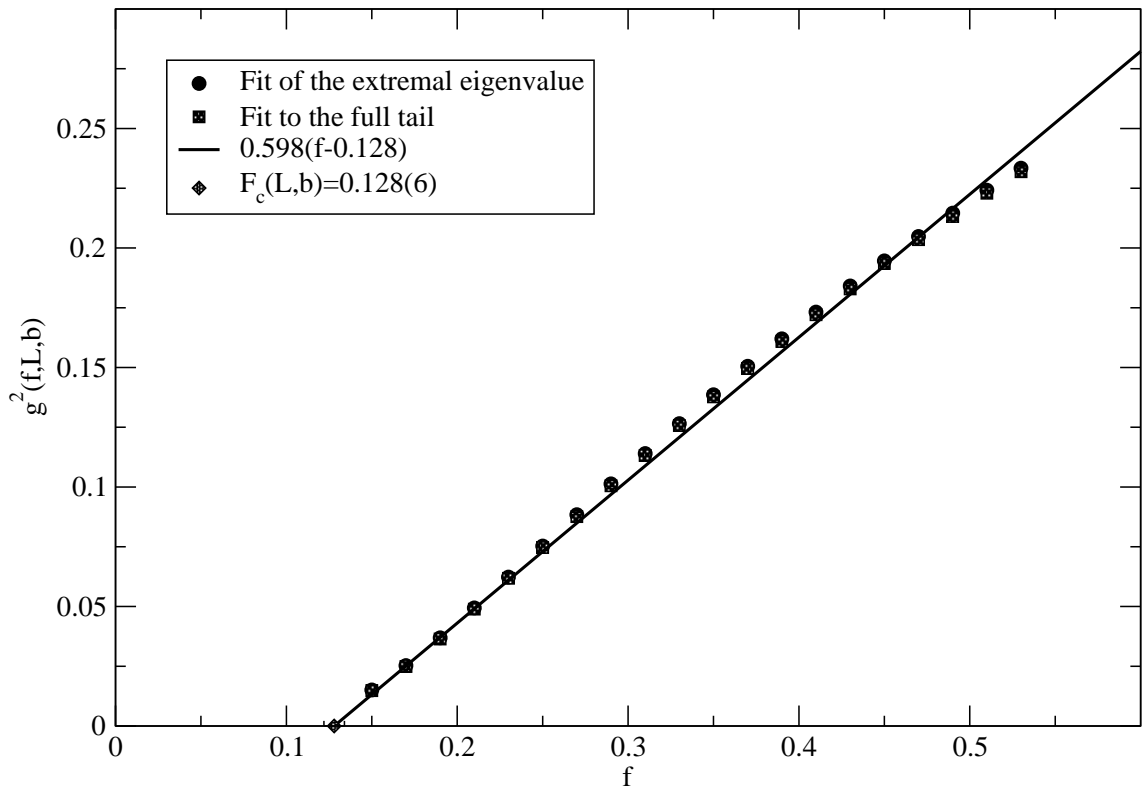
\caption{ Plot of
$g^2(f,L,b)$ as a function of $f$ at $b=0.361653$, $N=44$ and
$V=7^4$. A fit to the data along with the extrapolated value for
$F_c(L,b)$ is also shown. }
\label{gap_vs_ff}}

\subsubsection{The continuum limit of $F_c(L,b)$.}

We now proceed to take the continuum limit of $F_c(L,b)$.
We used two loop tadpole improved perturbation theory
to keep the physical loop fixed as we varied $b$.
Using~\cite{knn}, we have a numerical formula for the critical
deconfinement temperature in lattice units, in terms of tadpole
improved perturbation theory, with $e(b)$ being the average value of the
plaquette for untwisted single plaquette gauge action at coupling
$b$.
\be
\frac{1}{T_c(b)} = 0.26
\left[\frac{11}{48\pi^2 b_I}\right]^{\frac{51}{121}}
e^{\frac{24\pi^2 b_I}{11}};\ \ \ \
b_I=be(b)
\label{twoloop}
\ee
This will be used to determine the lattice spacing in physical units.

We set $lt_c=LT_c(b)$ and accumulate data at several values of the
physical loop size: $lt_c=0.42,0.48,0.58,0.64$. At each of these
physical sizes, we went through a sequence of three to four
lattice spacings $a(b)$, and, for each $b$, we obtained a value
for $F_c(L=\frac{l}{a(b)},b)$, using the methods explained above.

For each $lt_c$, the $F_c(L=\frac{l}{a(b)},b)$ values were then
extrapolated to the continuum assuming
\be
F_c(L=\frac{l}{a(b)},b) = f_c(lt_c) + O(T_c^2(b)).
\ee
where we recall
that $T_c(b)=a(b)t_c$. (We shall use the slightly inconsistent
notations $f_c(l)$ and $f_c(lt_c)$ for the same quantity, where the
argument is either dimensional or made dimensionless by the appropriate
power of the physical deconfining temperature $t_c$.)

The extrapolated results for $f_c(lt_c)$ are
shown in figure~\ref{crit2}.

\FIGURE[ht]{
\epsfig{file=crit2.eps,  width=\textwidth}
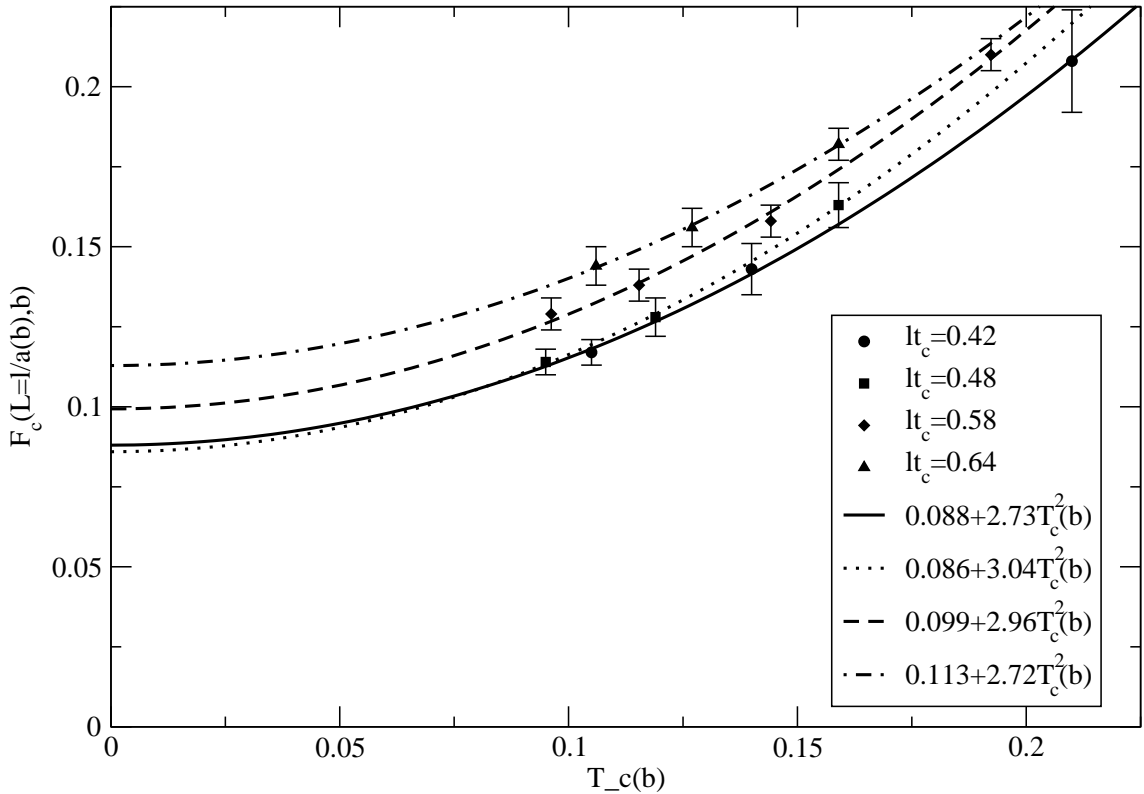
\caption{ Plot of $F_c(L=\frac{l}{a(b)},b)$ as a function of $T_c(b)$ for
four different values of $lt_c$. Also shown are fits to the data
producing continuum values for $f_c(lt_c)$. }
\label{crit2}}

Our final results are plotted in the $(f,l)$ plane. The critical
line $f_c(l)\equiv f_c (lt_c)$ divides this plane into two as
shown in figure~\ref{fltc}. In the upper part Wilson loops have a
gap in their eigenvalue distribution, whereas in the lower part
there is no gap. The band bounded from below
by the critical zero-lattice-spacing line indicates the range
of finite-lattice-spacing effects.

\subsection{A study of both sides of the transition under scaling.}

Finite lattice size effects are evident in figure~\ref{fltc}. It
is impractical to just scale the loop because the curve $f_c(l)$
varies when $l$ is changed in the appropriate range by less than
the lattice corrections. To see the effect of scaling we need to
go on a line in the $(f,l)$ plane that is not horizontal, which
means we are going to vary the smearing as a function of the
physical loop size. As usual, we trade $l$ for the dimensionless 
variable $lt_c$.

We chose the line (we shall refer to this line as the ``crossing line'')
\be
f = 0.25 - (0.6128 lt_c)^2
\label{fline}
\ee
shown in figure~\ref{fltc} as it cuts across the entire finite band of lattice
corrections bounded by the critical line $f_c(lt_c)$.

Consider a fixed loop of size $L$ on the lattice for various
lattice couplings $b$ such that $f(b)$ is given by (\ref{fline})
with $lt_c=LT_c(b)$ and $T_c(b)$ given by (\ref{twoloop}). We wish
to study the nature of the transition in the eigenvalue
distribution along this line.

\FIGURE[ht]{
\epsfig{file=fltc.eps,  width=\textwidth}
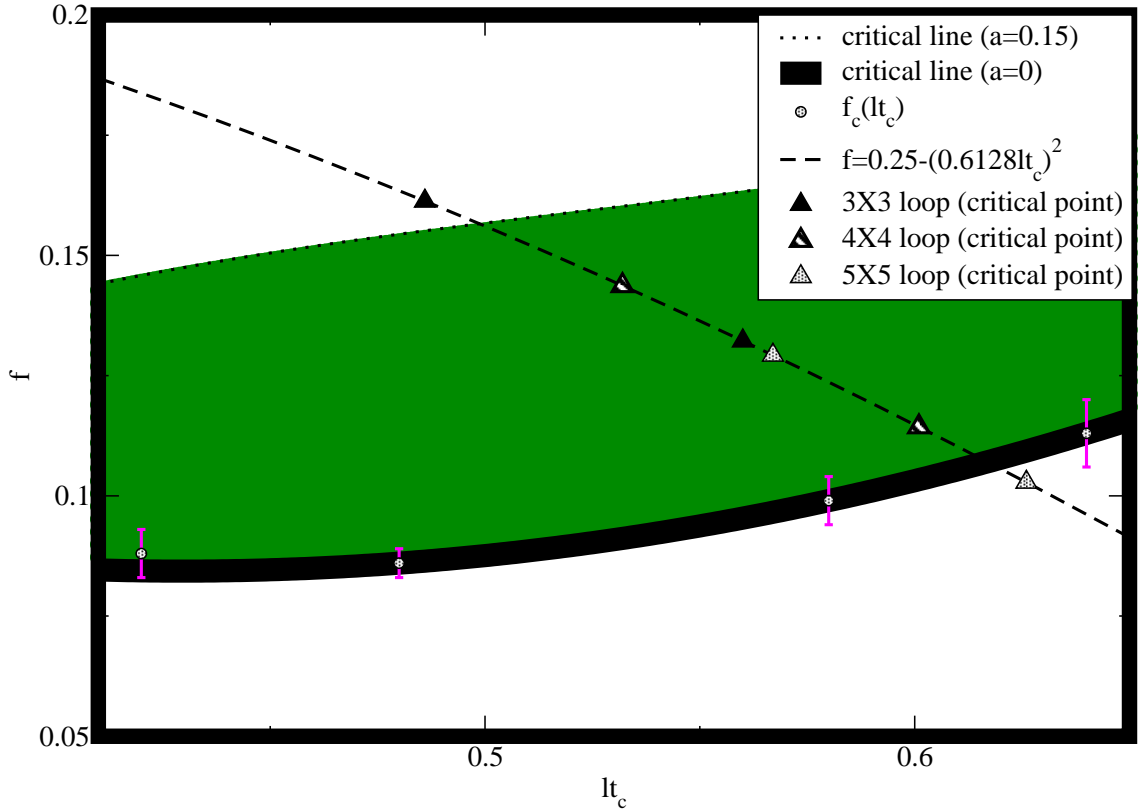
\caption{ The critical line in the
$(f,lt_c)$ plane. Also shown is the crossing line used to
investigate the phase transition induced by scaling the Wilson
loops. }
\label{fltc}}

We already know that the distribution should have no gap for $l
> l_c$ whereas it should have a gap for $l < l_c $. In
addition to getting an estimate for $l_ct_c$, we can also
investigate the universality class of this transition by keeping
$L$ fixed and going along the crossing line given by
(\ref{fline}). We set $L=3,4,5$ and worked in the following ranges
of $b$: for L=3, in $0.345 < b < 0.365$, for L=4, in $0.346 < b <
0.368$ and for L=5, in $0.351 < b < 0.369$. In each range, as $b$
decreases the physical loop is scaled up. In this way one gets
overlaps at the same physical loop size, once represented by a
smaller value of $L$ and a smaller value of $b$, and other times,
by correspondingly larger values of these parameters. Because of
scaling violations one will not get identical eigenvalue
distributions; we need to see that, as $b$ and $L$ increase, these
corrections decrease and that a finite limiting eigenvalue
distribution is approached. This is the eigenvalue distribution
associated with the continuum loop of size $l$.

In Figure~\ref{fltc} we display, for each value of $L=3,4,5$, the
two critical lengths one obtains when extrapolating from each
phase independently as triangles on the crossing line. As $L$ is
increased, we see that the distance between the two members of
each pair decreases as $L$ increases. Also, the middle points of
the segments connecting the points in each pair appear to converge
as $L$ increases. This indicates that in the continuum limit there
will be one single transition point on this line, a loop size at
which a continuous transition in the large $N$ spectrum of ${\hat
W}$ occurs.

\subsubsection{Estimating the critical size from the small loop
side.}\label{gapside}

We again obtained estimates for the infinite $N$ gap, using the
extremal eigenvalue distribution ($p(s)$) and the eigenvalue density
($p_f(s)$) near
the edge. The estimates for the gap came out consistent with each other
and the results are plotted in figure~\ref{gap_diag}. The
critical size of the loop is extracted by again assuming
a square root singularity written in the form
\be
g^2(lt_c)= A \ln \frac{l}{l_c} \ee
For each value of $L=3,4,5$ we
got an estimate for $l_ct_c$ which is shown in
figure~\ref{gap_diag}. One sees that the lines $L=4,5$ are closer
to each other than the lines $L=3,4$; the ratio of the approximate
distances between the two pairs of lines is in rough agreement
with a hypothesis of order $a^2 (b)$ corrections.

\FIGURE[ht]{
\epsfig{file=gap_diag.eps,  width=\textwidth}
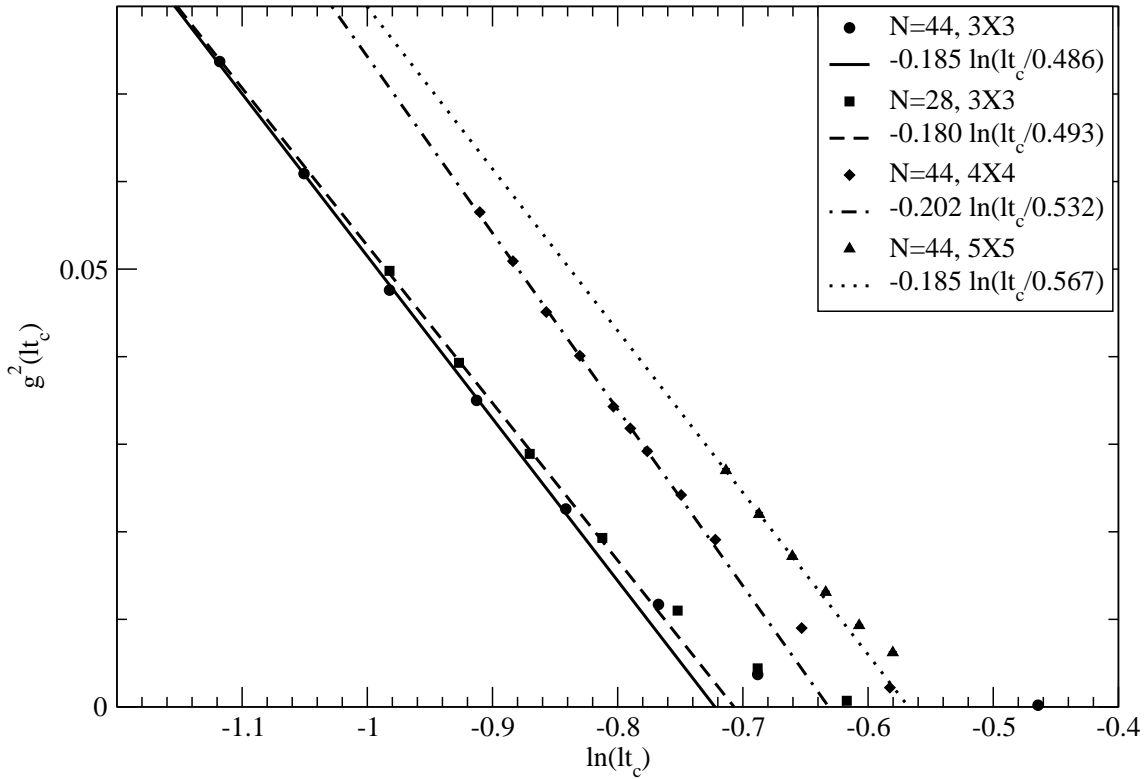
\caption{ The square of the
gap as a function of $\ln(lt_c)$ along the crossing line. }
\label{gap_diag}}

\subsubsection{Estimating the critical size from the large loop
side.}\label{rhopi}

For operators with no gap we need a method to extract the infinite
$N$ value of the eigenvalue density at angle $\pi$, $\hat\rho
(\pi)$. Shrinking the loop and/or smearing it more makes $\hat\rho
(\pi)$ decrease. We also need a method to extrapolate to the point
where $\hat\rho(\pi)$ vanishes, without getting too close to it.

The estimation of the infinite $N$ value of $\hat \rho(\pi)$ also
suffers from large subleading effects, of order $\frac{1}{N}$
typically. There is a rapid fluctuation in the eigenvalue density
reflective of the repulsion between eigenvalues. The rapid
fluctuation is eliminated by Fourier transforming the numerically
obtained binned eigenvalue distribution with respect to the angle
for wave-lengths kept large relative to $\frac{1}{N}$ . This is
equivalent to evaluating the traces of $\hat W^k$ for consecutive
values of $k$, all kept much smaller than $N$. Keeping only those
Fourier modes, we get our estimates for $\hat \rho(\pi)$; checks
against a simulation at lower $N$'s shows that we are determining
$\hat \rho(\pi)$ to reasonable accuracy.

The extrapolation of $\hat \rho(\pi)$ to zero as the loop size is
decreased and/or its amount of smearing is increased is done in a
similar way as when we deal with the gap. We find that the leading
behavior is again square-root like, but this holds in a small
range which can be extended by adding more powers of $(l-l_c)$.
The determination of $l_c$ by this method is seen to be quite
robust.

In two dimensions the large $N$ eigenvalue distribution for
loops of arbitrary size is known in the continuum \cite{duol}.
As we shall elaborate later, it is plausible to assume
that the behavior close to the large $N$
transition is universal, and holds also for
our $\hat \rho(\theta)$ in four dimensions.
Therefore, we
postulate that
${\hat \rho(\pi)}$
vanishes as a square root
at the critical value of any parameter
that enters the
underlying distribution controlling
the smeared Wilson loop operators in an
analytic manner. This is why we took $\hat\rho (\pi)$ to vanish as
$\sqrt{l-l_c}$ as one approaches
the appropriate critical size, $l_c$, from above.

As indicated above, for $l$ large enough with respect to $l_c$
we extract the nonzero value of $\hat\rho(\pi)$ by first
representing the distribution for all angles as:
\be
\hat\rho(\theta) = \sum_{k=0}^N f_k \cos(k\theta)
\ee
Next, we truncate the Fourier series and use the truncated series
to provide the value of $\hat\rho(\pi)$.
\be
\hat\rho(\pi) = \sum_{k=0}^{k_m} (-1)^k f_k
\ee
The value of $k_m$ is
chosen to be slightly less than $N/2$; this eliminates
oscillations of frequency $N$. We checked that our choice for
$k_m$ was reasonable by looking for stability as we vary $k_m$.

We
found that for low enough $k$'s $f_k$ oscillates in sign from
$k=1$ onwards.
As Figure~\ref{fourier} shows, for $k$ up to about $\frac{N}{2}$
the coefficients are smooth in $k$, and the sign oscillation is
obeyed. We shall come back to this property later.

\FIGURE[ht]{
\epsfig{file=fourier.eps,  width=\textwidth}
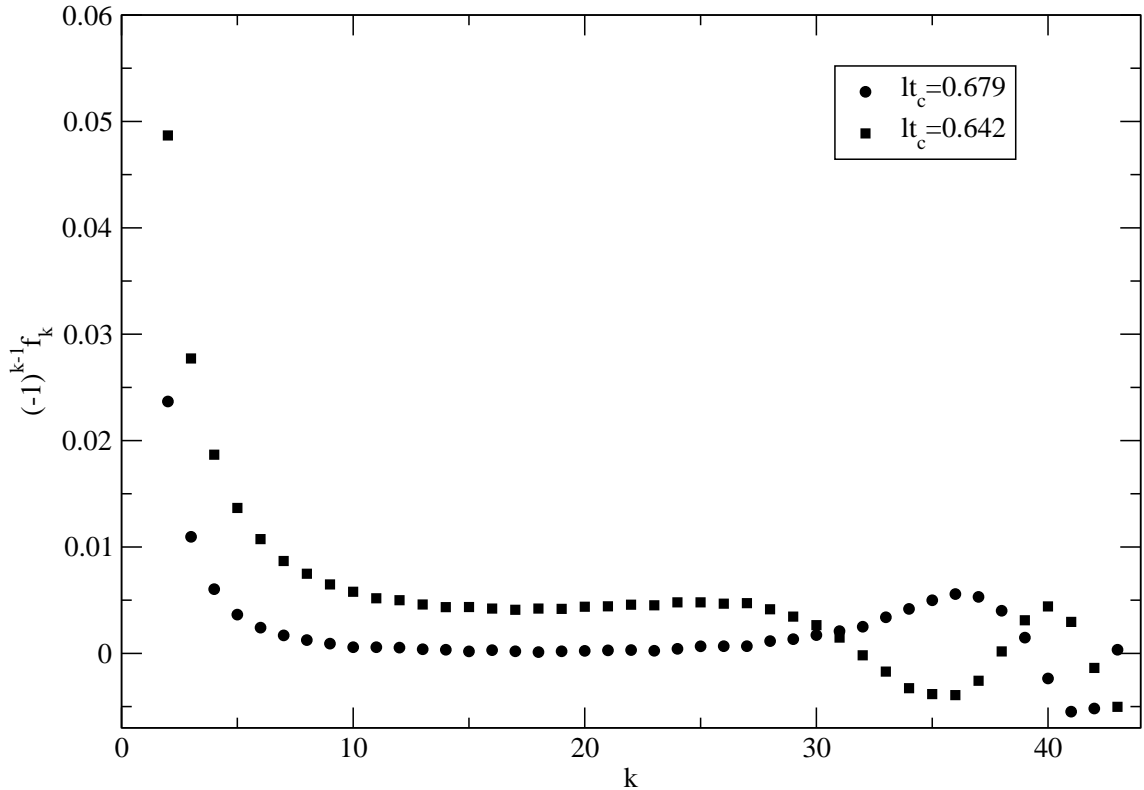
\caption{ The fourier
coefficients $f_k$ as a function of $k$ for two different loops in
the gapless region. The measurements were obtained using an $L=5$
loop at $N=44$ at two different lattice spacings. }
\label{fourier}}

The values of $\hat\rho(\pi)$ we obtain are fitted to a square root
in the critical region and the results are shown in figure~\ref{rho_diag}.
The estimates for $l_ct_c$ from $\hat\rho(\pi)$ tend to be higher
than the ones coming from fitting the gap. This difference
decreases as the spacing goes to zero.

The critical values for $l_ct_c$ obtained from the gap and
$\hat\rho(\pi)$ are shown in figure~\ref{fltc}. One can see that
the difference between the two estimates shrinks and one can also
see that the average of the two estimates approach a finite limit
as the lattice spacing decreases.

The size and behavior of the
scaling violations we see are consistent with usual expectations,
reflecting contributions
of higher dimensional continuum operators to the action and the
operators. We leave for the future a more detailed investigation of
these scaling violations, but note that the orders of magnitude
are typical.

\FIGURE[ht]{
\epsfig{file=rho_diag.eps,  width=\textwidth}
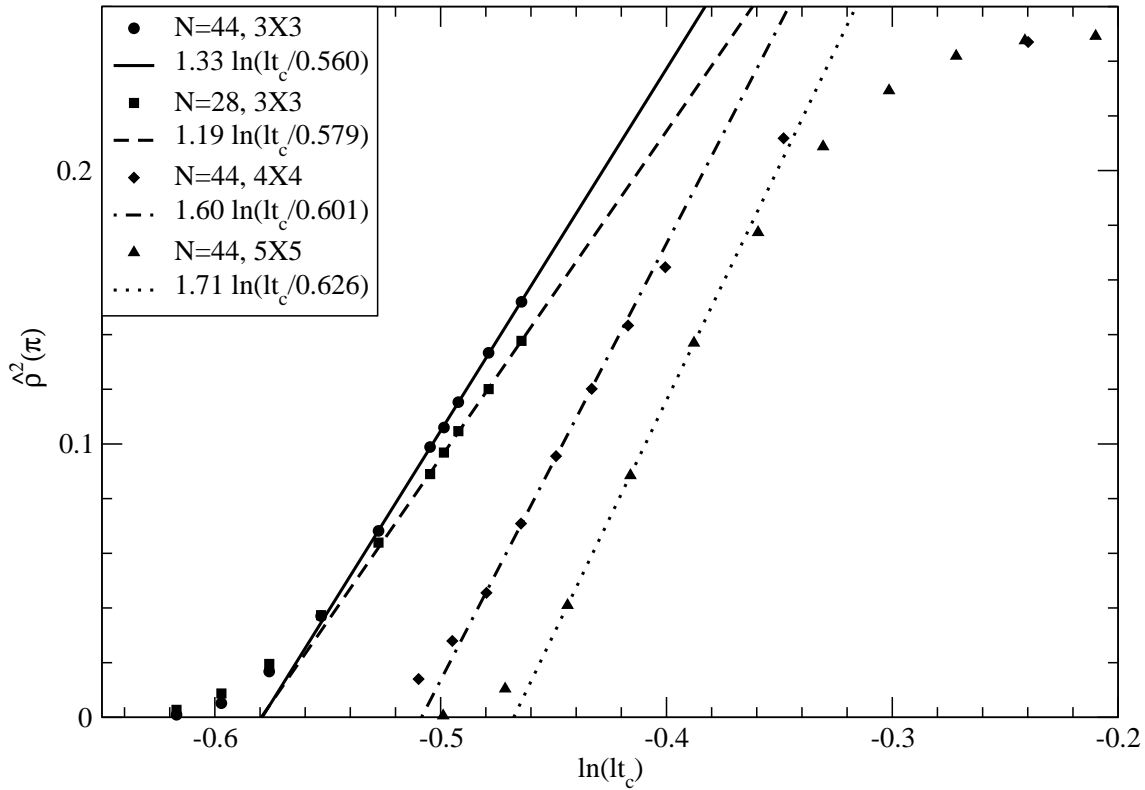
\caption{ The square of the
$\hat\rho(\pi)$ as a function of $\ln(lt_c)$ along the crossing line. }
\label{rho_diag}}

\subsection{General features of $f_c (l)$.}

In summary, using all of our data, a rough estimate for the
critical size on our crossing line in the $(f,l)$ plane, is $l_ct_c\approx
0.6$. $l_c\approx \sqrt{\sigma}$, where $\sigma$ is the string
tension and we used the approximate relation $0.6 \sigma=t_c$. In
QCD units, $l_c$ is about half a fermi, in general accordance with the
accumulating body of numerical data which sets the transition from
perturbative to string-like behavior at rather short scales, of
less than one fermi. Clearly, varying the crossing line, the loop shape
and/or the details of the continuum construction of the Wilson
loop operator will produce slightly different values of $l_c$.
However, the small variation over scales in the critical line
$f_c( l)$ we find indicates that the transitions will occur in a
restricted range of scales. Numerically, it seems that $f_c(l)$ asymptotes
to a positive constant as $l\to 0$. This excludes choices of
Wilson loop definitions with $l_c t_c << 1$. However, if it is 
somehow determined by
further simulations on very fine lattices that $f_c(l)\to 0$
as $l\to 0$, such choices must be recognized as possible;
in that case it is unlikely that an effective string description will
hold for $l$ too close to $l_c$. Even if
it were possible, we would not choose such a definition of regulated 
continuum Wilson loop operators.

We have seen the scaling violations in the $F_c(L,b)$
approximations to $f_c (l)$ in a range of physical scales close to
the large $N$ transition point. It is impossible to decrease the
physical size of the loops much more, because that would require
too large
values of $b$ for the large $N$ twisted reduction trick
to work on a $7^4$ lattice. Similarly, it is impossible to get to
much larger loops because working at too small values of $b$ takes
the lattice system through a first order phase transition to a
strong coupling phase which has no relevance to continuum physics.

We found that
$f_c(l)$ flattened out for small loops. It is possible that
$f(l)$ does decrease to zero as $l$ goes to zero, but if this at
all happens, one would expect the effect to be logarithmic (i.e.
$f(l) \sim\frac{1}{|\log \Lambda l|}$), since $f(l)$ ought to
behave roughly as the gauge coupling constant at scale $l$, which
enters the perimeter divergence smearing is intended to eliminate.
When $l$ increases, $f_c(l)$ starts increasing more rapidly and
eventually, $f_c(l)$ would violate its lattice bound. It is a
strain to arrange for too large loops to have a spectral gap,
because for large loops confinement is reflected by an almost
uniform eigenvalue density.

Note however that confinement is not necessary in order to have
our large $N$ transitions; even if the eigenvalue density does not
approach a uniform distribution for very large loops it remains
possible that large loops have no gap in their spectrum. Thus, a
similar analysis might apply also to models that do not confine.

\section{The continuum view.}

We would like to formulate everything we have been doing in
continuum language because the basic phenomenon is, we claim, not
a lattice artifact. So, we need to understand in continuum
language what smearing means, what the smeared observables are,
and how one could calculate them using perturbative and
non-perturbative continuum methods. Next, we would need at least
to consider how the large $N$ transitions would fit into a
continuum description and whether they are at all acceptable in a
unitary field theory.

\subsection{Smeared continuum Wilson loops and hints of a fifth dimension.}

Our first task is to understand what smearing means in the
continuum. The answer is that the quantity $f n$ plays the role of
the fifth ``time'', $\tau$, in the Langevin equation when the
noise has been set to zero:
\be
\frac{\partial A_\mu}{\partial\tau} =D_\mu F_{\mu\nu}
\label{langev}
\ee
This can be seen by comparing equations (2.4-2.7) to (4.1) for
small $f n$ and small $\tau$. 

If we add a Zwanziger term \cite{zwa135} to the right hand side
(it has no effect on our Wilson loops or on any other gauge
invariant observable) and move it to the left hand side, the
equation of motion gets a five dimensional form, with $x_5\equiv
\tau$:

\be
F_{5\nu} =D_\mu F_{\mu\nu} \ee

When a noise term is added to the right hand side one gets the
Parisi-Wu stochastic quantization of gauge theories \cite{stoch13}.
This setup has been shown to correspond to a topological, gauge
invariant five dimensional theory theory \cite{balzwa14}, in which
the fifth direction plays a special role (for example its units
are length square as opposed to the units of the other four
coordinates). There is Euclidean rotational invariance only in four
dimensions. The bulk theory is trivial in the sense of a
topological field theory, but the boundary has nontrivial
dynamics. This setup is reminiscent of the concept of holography.
The noise term is an essential ingredient in the construction, and
it is an open and perhaps interesting question whether our work
can be made to have a more precise relation to the one of \cite{balzwa14}.

In our paper we only looked at loops at one common fixed $\tau$.
This $\tau$ determined the thickness of the loop which we took as
uniform. But, it makes perfect sense to also consider loop
operators where $\tau$ varies round the loop, describing a string
of varying thickness. Moreover, one would have to consider such
loops if one looks for a loop equation that is local all around
the loop. Indeed, the ordinary four dimensional equation of motion
enters the $\tau$ evolution and therefore it is possible that at
the expense of considering loops of every
possible thickness simultaneously one can
restore the locality of the Migdal-Makeenko equation that we lost
when we simply fattened the loops staying in four dimensions.

In turn the Migdal-Makeenko equations are perhaps the right
starting point to get at the fundamental (infinitely thin) large
$N$ QCD string, which would live now in five warped dimensions and
might provide a description of four dimensional physics that is
equivalent to that provided by four dimensional fat strings. This
speculation suggests
that our simple objective to
get well defined continuum Wilson loops as unitary matrices in
four dimensions
did not lead us towards and extra dimension
by sheer coincidence.

\subsection{Instantons close the gap.}

For a small loop the parallel transporter round it ought to be
close to the identity matrix and have no eigenvalues around $-1$.
At infinite $N$, one expects a total suppression of eigenvalue far
from $1$, and a large gap. In matrix models, for finite but large
$N$, this suppression is typically exponentially small in $N$.

Consider a circle of radius $R$ in a plane and an instanton
with center placed somewhere on the circumference of the circle.
We can easily calculate the eigenvalues of the Wilson loop
operator for the circle, opening it at the instanton center.

The instanton gauge field in a regular gauge is given by:
\be
A_\mu=\imath\frac{x^2}{x^2+\rho^2} U^\dagger \partial_\mu U,
\ee
where
\be
U=\frac{x_4+\imath\vec x\cdot\vec\sigma}{\sqrt{x^2}}.
\ee

Defining our circle as $(x_4-R)^2+x_3^2=R^2$, with $x_{1,2}=0$, we find:
\be
{\cal P} e^{\imath\oint A_\mu dx_\mu}=e^{\imath\pi
(1-\frac{\rho}{\sqrt{\rho^2+4R^2}})\sigma_3 }  \ee

So long as the instanton is small enough and close enough to the
periphery of the circle it would create some eigenvalue density at
$-1$. Large instantons are irrelevant for $\Lambda R <<1$, so one
does not need to face the well known
infrared problem of instanton calculus if the loop is small. This is
a $e^{-N\cdot {\rm Const}}$ effect, likely to join smoothly
onto the asymptotic behavior we would get from any matrix-model
universal (double-) scaling function. This might be enough to select
a unique solution to the differential equation of the matrix model
and the latter may take us all the way to large $R$, where an
effective string theory has an overlapping regime with the matrix
model. Clearly, without some explicit calculations this is just a
suggestion.

Note that the continuum version of smearing, as we defined it, will
not affect the instanton since it sets to zero the right hand side
of~(\ref{langev}).

\subsection{A guess for the large $N$ universality class.}

There are at least two obvious candidates for the universality
class of the large $N$ transitions in our smeared Wilson loops. If
one just asks for the generic model dealing with a unitary matrix
that opens a gap in its spectrum and which obeys a reality
condition reflecting CP invariance, one will end up with a single
unitary matrix model and the famous Gross-Witten~\cite{gw12} large $N$
transition. The other immediate option is two dimensional QCD
defined on a sphere of dimensionless area $A$ which has the
Douglas-Kazakov transition. The latter transition also occurs for
two dimensional QCD on the infinite plane, but not for the
partition function this time, but, rather for Wilson loop
operators of area $A$; the transition on the sphere in one
case and in observables in the other are mathematically identical.
We now argue that the second option is preferred, both
theoretically and numerically.

It may be a bit confusing that we distinguish these two cases, as
both are relevant to two dimensional QCD. The Gross-Witten
transition is relevant to lattice QCD with a single plaquette
action of the Wilson type.
The single
unitary matrix in the model is the parallel transporter round an
elementary plaquette. As a function of the lattice coupling,
this matrix opens a gap at
some critical value as $b$ is increased from zero and the
trace of this matrix will be nonanalytic at the transition point.
Now, consider the lattice model for which the
analysis of Gross and Witten applies, but consider a loop
larger than $1\times 1$. It is still true that its trace will be
nonanalytic at the Gross-Witten transition point. But,
the parallel transporter
round the larger loop will open a spectral gap at a
large value of $b$, and at that value the trace of the larger loop
does not exhibit a nonanalytic behavior. In short, the nonanalytic
behavior of the trace of the plaquette is a lattice effect which
has no relevance to the continuum limit. Once we agree that we
should avoid looking at lattice effects, it is obvious we should
rather consider continuum two dimensional QCD as providing a
representative of the large $N$ universality class relevant to the
transition we found for our smeared four dimensional Wilson loops
${\hat W}$.

In continuum two dimensional planar QCD the spectrum of Wilson
loop operators, for smooth simple loops, does not depend on the
the shape of the curve but only on the enclosed area. As the loop
is scaled, all that matters is that the area changes, and at
infinite $N$ it is known that its eigenvalue distribution
undergoes a transition at a specific area (measured in units of
the two dimensional gauge coupling constant) where it just opens a
gap at -1. However, the traces of any finite power of the Wilson
loop operator stay analytic as the area goes through the critical
point of the spectrum. The expectation value of an $n$-wound
Wilson loop in two dimensions still depends just on its
fundamental area, which we make dimensionless using the 't Hooft
coupling, and denote by $A$. The explicit formula at large $N$ is
known \cite{KKwloop145}, given by the product of a polynomial and
an exponential:
\be
\frac{1}{N}\langle Tr { W^n } \rangle = \frac{1}{n} L_{n-1}^{(1)}
( 2A n) e^{-An}
\label{twodwil}
\ee
The dependence on scale hardly could be more analytic.

The Gross-Witten universality class and the one associated with
continuum QCD are related but not necessarily identical. Still,
both candidates produce a Painleve II equation in the double
scaling limit \cite{peris, gross-mat}.

Another point to keep in mind is that the string theory associated
with the universal class of the Gross-Witten transition is quite
degenerate and appears to differ significantly from any
putative string theory dual of QCD \cite{ias-double-cut}. 
On the other hand, there does
exist a more tangible string description of two dimensional QCD
due to Gross and Taylor \cite{grosstay15} which also has sigma
model representations \cite{moorhorav155}. The large $N$
transition in Wilson loops of continuum two dimensional QCD on
infinite space-time is of the same type as the Douglas-Kazakov
transition in the partition function of planar two dimensional QCD
defined on a sphere of area $A$ \cite{dougkaz157}.

There is another feature in favor
of two dimensional QCD as the universality
class and it has to do with instantons. When one takes the
nonabelian instanton solution that we showed makes a non-zero
contribution to the eigenvalue density at $-1$ for small loops in
four dimensions and looks only at a two dimensional slice
one gets two dimensional instantons that were
argued to be related to the Douglas-Kazakov transition
in~\cite{gross-mat}.

All this adds up to putting the Gross-Witten transition in
disfavor as a representative of the universality class of our
transitions, and leaves us with the preferred option of two
dimensional planar QCD. An explicit formula for the
eigenvalue distribution of Wilson loops can be found in~\cite{duol}
and it is possible~\cite{inst2d}
to connect this formula to (\ref{twodwil}).

\FIGURE[ht]{
\epsfig{file=durolefit.eps,  width=\textwidth}
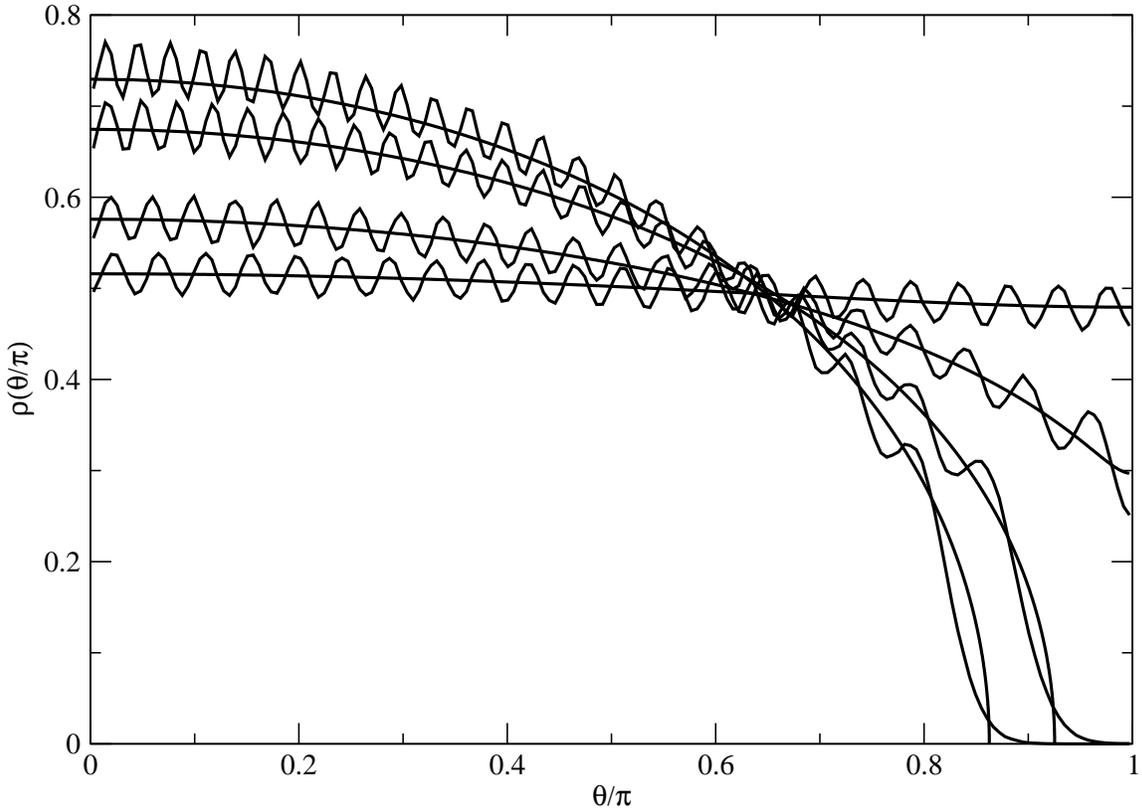
\caption{ Fit of the
distributions to the Durhuus-Olesen distributions for four
different sizes of Wilson loops, namely,
$lt_c=0.740,0.660,0.560,0.503$. The associated areas ($k$ in the
Durhuus-Olesen notation) that describe the continuous curves are
given by $k=4.03,2.30,1.41,1.15$ respectively. }
\label{durolefit}}

To see that we are not completely off in our guess we show a plot
of several eigenvalue distributions we measure
versus the exact infinite $N$
continuum solution of Durhuus and Olesen \cite{duol}:

Figure~\ref{durolefit} shows shows how the critical size is
traversed and one sees directly the effect of the transition and
that the Durhuus-Olesen distribution seems to work pretty well on
both sides of the transition. An attempt to fit to the
Gross-Witten distribution would fail, since on the gap-less side
the Gross-Witten distribution would always cross the value
$\frac{1}{2}$ in Figure~\ref{durolefit}
at $\theta=\pm\frac{\pi}{2}$.

Another point in favor of the universality class is evident from
equation~\ref{twodwil}: The prefactor has oscillating signs for
large areas. These are nothing but the oscillating signs we saw in
the Fourier coefficients of Figure~\ref{fourier}. These
oscillating signs could be taken as an indication that an
effective string representation of loops traversed multiple times
requires some elementary two dimensional excitation of statistics
different from that of bosons~\cite{gross-ooguri}. The simple
effective string models one typically uses do not address loops
traversed multiple numbers of times.

Because of the importance of the large $N$ universality class to our
program, we now turn to a more general description of what appears to
be its fundamental representative family in terms of matrix models:

Imagine $n$ $SU(N)$ matrices, independently and identically
distributed according to a conjugation invariant measure
$d\mu(U_i)$. The measure is peaked at $U_i=1$ and is of the single
trace type. As a function of each eigenvalue it decreases
monotonically as one goes round the circle from $1$ to $-1$, where
it has a minimum. The continuum Wilson loop operators, $\hat W$,
one defines, would behave in the vicinity of the large $N$
transition point at which their eigenvalue distribution opens a
gap at $-1$, similarly to $W=\prod_i^n U_i$. The parameters of the measure
and $n$ can be adjusted to various critical behaviors in the
eigenvalue distribution of $W$ at infinite $N$, and likely produce
a variety of double-scaling limits corresponding to large $N$
fixed points of decreasing degrees of stability. The shape and
scale of ${\hat W}$ map analytically into these parameters in the
vicinity of the transition. Most plausibly, the most generic and
stable transition is the relevant one.

This random matrix model is exactly soluble, as we now sketch,
leaving details for the future:

One introduces $n$ pairs of $N$-component Grassmann variables,
$\bar\psi_i, \psi_i$ and proves by simple recursion that
\be
\int \prod_i^n ( d\bar\psi_i d\psi_i ) e^{z\sum_i^n
\bar\psi_i\psi_i - \sum_i^n \bar\psi_i U_i\psi_{i+1}}=\det(z^n + W )
\ee
Above, we use the convention $\psi_{n+1}\equiv - \psi_1$.

Now, one can introduce auxiliary
fields for the $SU(N)$ invariant bilinears
$\bar\psi_i \psi_i$ and $\bar\psi_{i}\psi_{i+1}$. Averaging over
the $U_i$ variables can be done independently and the remaining
integral can be dealt with at infinite $N$ by ordinary saddle
point methods.

Choosing an axial gauge on an infinite two dimensional lattice, it
is easy to see that both on the lattice and in the continuum
\cite{duol} the non-minimal Wilson loops of two dimensional planar
QCD are indeed described by the above universality class.
Similarly, the Douglas-Kazakov critical point is also in this
universality class, corresponding to the most stable critical
point.

Intuitively, the universality class can be visualized as a
nonabelian generalization of the following abelian situation:
Let there be given a smooth loop $C$ and let $S$ be a surface
of the topology of a disk bounded by $C$. On the surface $S$
define a two dimensional  $U(1)$ gauge field, with independently
and identically distributed fluxes through small patches of
surface that tile  $S$. The phase of the parallel transport
round $C$ is then given by the sum of all these fluxes. The
distribution of that sum will be generically governed by the
central limit theorem. It is now obvious that we have a nonabelian
generalization of this arrangement~\cite{janik}, of the type studied by
Voiculescu \cite{voicu}.

\section{Summary and Discussion.}

We first give an overview of various ideas that have influenced us
and to which we have not yet referred directly. Subsequently we
proceed to a summary explaining what are the essential new points
we have made in this paper. We finish with a short description of
planned future work.

\subsection{Connections to other work.}

At the conceptual level, this paper has been influenced by the
work of many others, in particular people in our own subfield of
lattice field theory, even if this was not mentioned at any point
in our presentation until now. While it is impossible to avoid
omissions, we try here to draw attention to what stands out in our
memory at this point.

That an extra dimension might be needed for a local string dual of
the field theoretical description clearly is an idea with many
sources. The relationship between this extra dimension and a space
of eigenvalues has been emphasized by many, for example in papers
by A. Jevicki \cite{jevi}. The essential point is that for a
hermitian or unitary matrix the eigenvalues are restricted to a
one dimensional line and repel each other; as such they make up
dynamically, at infinite $N$, a continuous one dimensional degree
of freedom. In the context of matrix models this produces an
analogue of the extra dimension we have seen in higher dimensional
setups.

Using the lattice to make progress on the question of a possible
string dual of nonabelian YM theory has preoccupied a sizable
group of researchers. This new wave of lattice work on the large $N$
limit has been started by M. Teper, and in various collaborations
he has addressed many of the questions we address
here \cite{teper}. In particular, the question of the cross-over
has been studied in various contexts \cite{teper-cross-over}. There
is an interesting issue raised there suggesting a qualitative
difference between three dimensional and four dimensional YM gauge
theories with respect to observable dependent large $N$
transitions. The immediate implication is that it would be
interesting to repeat our work here in three Euclidean dimensions;
it would be surprising to see three dimensions so different in
this context from both two and four dimensions.

J. Kuti \cite{kuti} and collaborators have studied in great detail the cross-over
from field theoretical to string-like behavior and have
shown how various
string states reorganize themselves as one goes from short to long
QCD strings. The philosophy of matching various field theories
with known or suspected string-like excitations onto an effective
string theory at long distances underlies much of this work. This
work has not yet been extended to large values of $N$, but this is
recognized as a possible valid avenue for further steps in this
program.

We have been influenced significantly by papers by R. Brower
\cite{brower}. On the one hand side he and collaborators have
provided additional examples in which one can see the extra
dimension being connected to standard scale decompositions of four
dimensional processes of traditional interest. This interpretation
is very natural when one thinks in terms of a warp factor in the
context of the famous AdS/CFT correspondence in the conformal and
related cases. It provides a most convincing resolution of the old
puzzle of how a string theory can embody hard, field theoretic
ultraviolet behavior.

It has also been stressed by Brower that one should be more
ambitious than merely connecting onto an effective long distance
string theory, and, that the extra dimension could provide the
means to construct a complete stringy dual to YM, valid at all
distances, albeit too difficult for actual calculation at short
distances. It is with this in mind that although our strings are
fat, we maintain the zigzag symmetry, which would be a potential
clue about the infinitely thin string dual. Also, with this in
mind, we insisted on finding observables that have a continuum
field theoretical semiclassical expansion at small sizes, where
the effects of short distance fluctuations reveal themselves.

The effective string description has been studied in various
guises also by others \cite{other}, in addition to Polchinski
and Strominger; the approaches focused both on
open Dirichlet strings and on closed string wrapping round a large
compact direction of space-time.

Another related topic is the issue of ``Casimir scaling'', in the
context of the dependence of the string tension on the
representation of the source and sink the $SU(N)$ string connects.
It is known that ``Casimir scaling'' is exact when applied to
charges in the fundamental and in the adjoint representation
at $N=\infty$,
so long as we take $N\to\infty$ first and let the minimal area
spanned by the loop increase only after that~\cite{adlerneu}.
We wish to mention here a recent paper~\cite{negele} in the context
of ``Casimir scaling'' at finite $N$ where the
Wilson loop eigenvalue distribution has been studied on the
lattice for $SU(2)$.

Last but not least we wish to mention the old idea of
preconfinement \cite{amativ16}, in which one pictures a jet
producing process in QCD as evolving by first creating (in the
planar limit) finite energy clusters that already are color
neutral and only subsequently hadronize. This preparation for
hadronization is similar to the early signals of closing the gap
in the eigenvalue spectrum, before the density becomes uniform up
to corrections exponential in the minimal area spanned by the
loop, where one has full confinement. As mentioned previously, one
can have a large $N$ transition without ever making it all the way
to confinement.

\subsection{Main new ideas in this work.}

Our main change in vantage was a focus on the universal nature of
large $N$ transition which we hope to turn into a calculation that
can take us through the cross-over which connects the regime in
which field theory is the most convenient description to the
regime in which string theory is assumed to eventually become a
convenient description.

We see the beginnings of such a calculational scheme and feel that
were it successful, it would provide a concrete result on which to
build in the search for a more elegant and fundamental dual
representation of Yang Mills theory.

We found ourselves naturally led into something that looks like an
extra dimension and one that is intimately related to scaling at
that.

\subsection{Plans for the future.}

The first problem for the future is a more thorough investigation
of the critical regime of the large $N$ transitions and the
numerical identification of the right universality class. In
parallel an effort will be made to make the semiclassical
calculation based on instantons at the short distance end
concrete. The double scaling limits in the larger universality
class would need to be studied in detail; assuming that our guess
for this universality class is right, it should be possible to
construct those double scaling limits exactly.

The connection onto a convenient effective string model at the
strong coupling end is another direction of future activity; here,
the main tool would be numerical at first, and we expect to draw
from the experience of the many other workers on similar problems.
Our main focus will be on keeping $N$ large so the string coupling
constant can be set to zero as is often done even for $N=3$.

Also, we feel it might be useful to try to face the question
whether an effective string theory for planar QCD is indeed not
different from an effective string theory in a model which has
nothing to do with non-abelian gauge theory. For example, on the
face of it one would expect the zigzag symmetry to play no role at
leading order in the effective string theory, but maybe this is
not a correct assumption. An analogy is the chiral Lagrangian
describing pions in QCD, which indeed could come from a
fundamental theory that has no gauge fields. But, the complete
decoupling at all scales of all the mesons, well known to occur at
infinite $N$, is something the effective Lagrangian, by itself,
does not incorporate in a structural way; rather, many of its free
parameters need to be set to zero.

\acknowledgments

We acknowledge partial support by the NSF under
grant number PHY-0300065.
R.N. also acknowledges partial support from Jefferson
Lab. The Thomas Jefferson National Accelerator Facility
(Jefferson Lab) is operated by the Southeastern Universities Research
Association (SURA) under DOE contract DE-AC05-84ER40150.
H. N. also acknowledges partial support
by the DOE under grant number
DE-FG02-01ER41165 at Rutgers University.


\begin{thebibliography}{99}
\bibitem{old1}
H.~Neuberger,
  Phys.\ Lett.\ B {\bf 94}, 199 (1980);
H.~Neuberger,
  Nucl.\ Phys.\ B {\bf 179}, 253 (1981);
  H.~Neuberger,
  Nucl.\ Phys.\ B {\bf 340}, 703 (1990).
\bibitem{gw12}
D.~J.~Gross and E.~Witten,
  Phys.\ Rev.\ D {\bf 21}, 446 (1980).
\bibitem{bulk2}
R.~Narayanan and H.~Neuberger,
  Phys.\ Rev.\ Lett.\  {\bf 91}, 081601 (2003).
\bibitem{knn}
J.~Kiskis, R.~Narayanan and H.~Neuberger,
  Phys.\ Lett.\ B {\bf 574}, 65 (2003).
\bibitem{poly3}
A.~M.~Polyakov,
  Nucl.\ Phys.\ Proc.\ Suppl.\  {\bf 68}, 1 (1998).
\bibitem{malda4}
J.~M.~Maldacena,
  Adv.\ Theor.\ Math.\ Phys.\  {\bf 2}, 231 (1998)
  [Int.\ J.\ Theor.\ Phys.\  {\bf 38}, 1113 (1999)].
\bibitem{doublesca5}
D.~J.~Gross and A.~A.~Migdal,
  Phys.\ Rev.\ Lett.\  {\bf 64}, 127 (1990);
E.~Brezin and V.~A.~Kazakov,
  Phys.\ Lett.\ B {\bf 236}, 144 (1990);
  M.~R.~Douglas and S.~H.~Shenker,
  Nucl.\ Phys.\ B {\bf 335}, 635 (1990).
\bibitem{ps6}
 J.~Polchinski and A.~Strominger,
  Phys.\ Rev.\ Lett.\  {\bf 67}, 1681 (1991).
\bibitem{effft7}
  A.~V.~Manohar,
  arXiv:hep-ph/9606222.
\bibitem{cdg75}
C.~G.~.~Callan, R.~F.~Dashen and D.~J.~Gross,
  Phys.\ Rev.\ D {\bf 20}, 3279 (1979);
 C.~G.~.~Callan, R.~F.~Dashen and D.~J.~Gross,
  Phys.\ Rev.\ D {\bf 19}, 1826 (1979);
 C.~G.~.~Callan, R.~F.~Dashen and D.~J.~Gross,
  Phys.\ Rev.\ D {\bf 17}, 2717 (1978).
\bibitem{overlap}
H. Neuberger, Phys.\ Lett.\ B{\bf 417}, 141 (1998);
H. Neuberger, Phys.\ Lett.\ B{\bf 427},  353 (1998);
R. Narayanan, H. Neuberger,
Nucl.\ Phys.\ B{\bf 443}, 305 (1995).
R.~Narayanan and H.~Neuberger,
  Nucl.\ Phys.\ B {\bf 696}, 107 (2004).
\bibitem{twist8}
 A.~Gonzalez-Arroyo, R.~Narayanan and H.~Neuberger,
  Phys.\ Lett.\ B {\bf 631}, 133 (2005).
\bibitem{div9}
 V.~S.~Dotsenko and S.~N.~Vergeles,
  Nucl.\ Phys.\ B {\bf 169}, 527 (1980).
\bibitem{ape10}
T. DeGrand, Phys. Rev. D63 (2001) 034503;
M. Albanese et. al. [APE Collaboration], Phys.\ Lett.\ B{\bf 192}, (1987) 163;
M. Falcioni, M.L. Paciello, G. Parisi and B. Taglienti, Nucl. Phys. Nucl.\ Phys.\ {\bf
B251}, 624 (1985).
\bibitem{pertsmear11}
C.~W.~Bernard and T.~DeGrand,
  Nucl.\ Phys.\ Proc.\ Suppl.\  {\bf 83}, 845 (2000).
\bibitem{tracy} C. A. Tracy, H. Widom, arXiv:solv-int/9707001.
\bibitem{gff} T. M. Garoni, P. J. Forester, N. E. Frankel,
arXiv:math-ph/0504053.
\bibitem{duol}
B.~Durhuus and P.~Olesen,
  Nucl.\ Phys.\ B {\bf 184}, 461 (1981).

\bibitem{zwa135}
D.~Zwanziger,
  Nucl.\ Phys.\ B {\bf 192}, 259 (1981).

\bibitem{stoch13}
G.~Parisi and Y.~s.~Wu,
  Sci.\ Sin.\  {\bf 24}, 483 (1981);
P.~H.~Damgaard and H.~Huffel,
  Phys.\ Rept.\  {\bf 152}, 227 (1987).
\bibitem{balzwa14}
L.~Baulieu and D.~Zwanziger,
  Nucl.\ Phys.\ B {\bf 581}, 604 (2000).
\bibitem{dougkaz157}
M.~R.~Douglas and V.~A.~Kazakov,
  Phys.\ Lett.\ B {\bf 319}, 219 (1993).

\bibitem{KKwloop145}
V. A. Kazakov, I. K. Kostov, Nucl.\ Phys. B{\bf 176}, 199 (1980).

\bibitem{peris}
V.~Periwal and D.~Shevitz,
  Phys.\ Rev.\ Lett.\  {\bf 64}, 1326 (1990).
\bibitem{gross-mat}
  D.~J.~Gross and A.~Matytsin,
  Nucl.\ Phys.\ B {\bf 429}, 50 (1994).
\bibitem{ias-double-cut}
I.~R.~Klebanov, J.~Maldacena and N.~Seiberg,
  Commun.\ Math.\ Phys.\  {\bf 252}, 275 (2004).
\bibitem{grosstay15}
D.~J.~Gross and W.~I.~Taylor,
  Nucl.\ Phys.\ B {\bf 400}, 181 (1993).
\bibitem{moorhorav155}
P.~Horava,
  Nucl.\ Phys.\ B {\bf 463}, 238 (1996);
S.~Cordes, G.~W.~Moore and S.~Ramgoolam,
  Commun.\ Math.\ Phys.\  {\bf 185}, 543 (1997).
\bibitem{inst2d}
  A.~Bassetto, L.~Griguolo and F.~Vian,
  Nucl.\ Phys.\ B {\bf 559}, 563 (1999)
  [arXiv:hep-th/9906125].

\bibitem{gross-ooguri}
D.~J.~Gross and H.~Ooguri,
  Phys.\ Rev.\ D {\bf 58}, 106002 (1998).
\bibitem{janik} R. A. Janik, W. Wieczorek, J.\ Phys.\ A:\ Math.\ Gen.
{\bf 37}, 6521 (2004).
\bibitem{voicu}
D. V. Voiculescu, K. J. Dykema, A. Nica, ``Free Random Variables'' AMS | CRM,
(1992).
\bibitem{jevi}
S.~R.~Das and A.~Jevicki,
  Mod.\ Phys.\ Lett.\ A {\bf 5}, 1639 (1990).

\bibitem{teper}
F.~Bursa, M.~Teper and H.~Vairinhos,
  PoS {\bf LAT2005}, 282 (2005).
 M.~Teper,
  PoS {\bf LAT2005}, 256 (2005).
  M.~Teper,
  arXiv:hep-th/0412005.
\bibitem{teper-cross-over}
F.~Bursa and M.~Teper,
  arXiv:hep-th/0511081.

\bibitem{kuti}
J.~Kuti,
  PoS {\bf LAT2005}, 001 (2005).
\bibitem{brower}
  R.~C.~Brower,
  Acta Phys.\ Polon.\ B {\bf 34}, 5927 (2003);
    R.~C.~Brower and C.~I.~Tan,
  Nucl.\ Phys.\ Proc.\ Suppl.\  {\bf 119}, 938 (2003).
\bibitem{other}
M.~Luscher, K.~Symanzik and P.~Weisz,
  Nucl.\ Phys.\ B {\bf 173}, 365 (1980);
M.~Luscher and P.~Weisz,
  JHEP {\bf 0407}, 014 (2004);
M.~Caselle, R.~Fiore, F.~Gliozzi, M.~Hasenbusch and P.~Provero,
  Nucl.\ Phys.\ B {\bf 486}, 245 (1997);
J.~M.~Drummond,
  arXiv:hep-th/0411017.





\bibitem{adlerneu}
S.~L.~Adler and H.~Neuberger,
  Phys.\ Rev.\ D {\bf 27}, 1960 (1983).
\bibitem{negele}
  A.~M.~Brzoska, F.~Lenz, J.~W.~Negele and M.~Thies,
  Phys.\ Rev.\ D {\bf 71}, 034008 (2005).



\bibitem{amativ16}
D.~Amati and G.~Veneziano,
  Phys.\ Lett.\ B {\bf 83}, 87 (1979).




\end{thebibliography}
\end{document}